\documentclass[useAMS,usenatbib,onecolumn]{mn2e}
\usepackage{graphicx,amssymb,amsmath,times}

\title{On the contribution of nearby sources to the observed cosmic-ray nuclei}
\author[S. Thoudam ]{Satyendra Thoudam\thanks{E-mail: satyend@barc.gov.in}\\
Astrophysical Sciences Division,
Bhabha Atomic Research Centre, Trombay, Mumbai-400085,
Maharashtra, India}
\begin{document}
\date{}
\pagerange{}
\maketitle
\label{firstpage}
\begin{abstract}
The presence of nearby discrete cosmic-ray (CR) sources can lead to many interesting effects on the observed properties of CRs. In this paper, we study about the possible effects on the CR primary and secondary spectra and also the subsequent effects on the CR secondary-to-primary ratios. For the study, we assume that CRs undergo diffusive propagation in the Galaxy and we neglect the effect of convection, energy losses and reacceleration. In our model, we assume that there exists a uniform and continuous distribution of CR sources in the Galaxy generating a stationary CR background at the Earth. In addition, we also consider the existence of some nearby sources which inject CRs in a discrete space-time model. Assuming a constant CR source power throughout the Galaxy, our study has found that the presence of nearby supernova remnants (SNRs) produces noticeable variations in the primary fluxes mainly above $\sim 100$ GeV/n, if CRs are assumed to be released instantaneously after the supernova explosion. The variation reaches a value of $\sim 45\%$ at around $10^5$ GeV/n. Respect to earlier studies, the variation in the case of the secondaries is found to be almost negligible. We also discuss about the possible effects of the different particle release times from the SNRs. For the particle release time of $\sim 10^5$ yr, predicted by the diffusive shock acceleration theories in SNRs, we have found that the presence of the nearby SNRs hardly produces any significant effects on the CRs at the Earth.   
\end{abstract}

\section{Introduction}
Cosmic-rays (CRs) below the knee region ($\sim 3\times 10^{15}$ eV) are generally considered to be of Galactic origin. Though the nature of their sources are not exactly known, it is widely beleived that majority of them are accelerated in supernova remnant (SNR) shock waves. Such a hypothesis is mainly based on the similarity of the power supplied by a supernova explosion to the power required to maintain the CR energy density in the Galaxy ($\sim 10^{41}$ erg s$^{-1}$). In addition, studies using nonlinear effects produced by accelerated CRs have shown that the maximum energy of CRs that can be accelerated in SNRs is $\sim Z\times 10^{15}$ eV (where $Z$ is the total charge of the particle) which is close to the knee (Berezhko 1996).

It is quite expected that the presence of one or more nearby discrete sources can result in various significant effects on the observed properties of CRs. Studies concerning their contributions to the total CR electron flux can be found in earlier studies like  Mao $\&$ Shen (1972) and Cowsik $\&$ Lee (1979) under the framework of an energy independent CR diffusive propagation model and later, in Atoyan et al. 1995 and Kobayashi et al. 2004 assuming a more realistic energy dependent diffusion cofficients in the Galaxy. For CR nuclei, Lingenfelter (1969) studied the density variations at the Earth due to the local sources using a simple particle escape model and later, Erlykin $\&$ Wolfendale (2001) carried out Monte Carlo calculations assuming SNRs to be distributed stochastically throughout the nearby Galaxy. Strong $\&$ Moskalenko (2001) studied the influence of the discrete nature of SNRs on the CR proton densities in the Galaxy using their GALPROP CR propagation code. In addition, some recent studies tried to stretch the importance of local sources to explain some strongly observed features of CRs like the knee and the anisotropy. For instance, Erlykin $\&$ Wolfendale (2000) claimed that the knee in the CR spectrum can be attributed to the presence of a single recent supernova (as yet unidentified) in the local region.  Erlykin $\&$ Wolfendale (2006) further tried to explain the rise in the anisotropy amplitude as well as the change in its phase near the knee using a single source exploded in the direction from the Sun downward of the main CR flux, which comes predominantly from the inner Galaxy. Moreover, using a convection-diffusion CR propagation model, Thoudam (2006) tried to emphasise the necessity for the presence of at least one old nearby source in order to explain the observed proton flux below $\sim 100$ GeV/n. In another recent study, Thoudam (2007a) studied the effect of the nearby known SNRs on the observed CR anisotropy below the knee and later on, suggested the possibility of explaining the data above $100$ GeV/n by a single dominant source with properly chosen source parameters and claimed that the source may be an \textit{undetected} old SNR with a characteristic age of $\sim 1.5\times 10^5$ yr located at a distance of $\sim 0.57$ kpc from the Earth (Thoudam 2007b).

Taillet $\&$ Maurin (2003) pointed out that the majority of the CRs reaching the Earth are possibly emitted by sources located within a few kpc and hence, the propagation parameters that we usually derived from the observed secondary$/$primary (s$/$p) may only give a local information which may be different from other parts of the Galaxy. In another work, B$\mathrm{\ddot{u}}$sching et al. (2005) showed that even if the CR propagation parameters are assumed to be constant throughout the Galaxy, the discrete nature of the CR sources can produce a much larger fluctuations of the CR primary densities than that of the secondaries, implying a significant fluctuations in the s$/$p ratio. In this paper, we shall also study about the expected fluctuations on the CR primary and secondary spectra at the Earth, but by incorporating the nearby known CR sources (considered here as SNRs). Such an approach is expected to give a more detailed understanding about the CR density fluctuations and its implications as far as the position of the Earth is concerned.   

We plan the paper as follows. In section 2, we give a brief description about our model. In section 3, we calculate the CR primary and secondary spectra from a discrete point-like source and in section 4, we calculate for the CRs from a continuous and stationary source distribution. In section 5, we study the CR density variations at the Earth due to a single nearby source and in section 6, we give an application to the nearby SNRs and present a comparison of the calculated CR spectra with the observed data. Finally in section 7, we give a brief discussion about our results and its implications.

\section {Model description}
Our model assumes that a major fraction of the CRs detected at the Earth are liberated from sources which are distributed uniformly and continuously (both in space and time) in the Galaxy. We refer to these sources as the background sources and the CRs they emit as the background CRs. In addition, we also assume that there exists some nearby discrete sources which inject CRs in a discrete space-time model and whose contributions to the total CRs are yet to be investigated. The contributions of the discrete and the background sources will be treated seperately. CRs from the discrete sources will be discussed under the framework of a time dependent diffusive propagation model while those from the background sources in a steady state model. 

For the background CRs, the flattened shape of our Galaxy allows us to assume their diffusion region as a cylindrical disk of infinite radius with finite half-thickness $H$. Such an assumption of infinite radius is valid at least for CR studies at the position of the Earth, which is at a distance of $\sim 8.5$ kpc from the Galactic Centre, where the effect of the Galactic radial boundary (assumed to be $\gtrsim 20$ kpc) on the CR flux is expected to be negligible. This is because a substantial fraction of the CRs reaching the Earth are liberated from sources located within an approximate distance which is of the order of $H$ (see e.g. Taillet $\&$ Maurin 2003). The actual value of $H$ is not exactly known. The values estimated using different CR propagation models fall in the wide range of $(2-12)$ kpc (see e.g., Lukasiak et al. 1994, Webber $\&$ Soutoul 1998, Strong $\&$ Moskalenko 1998). For our calculations, we choose a value of $H=\pm5$ kpc. We further assume that the background sources as well as the interstellar matter are distributed in a thin disk of radius $R$ and half-thickness $h$. Here again, the effect of $R$ on the CR flux is negligible as long as $R>H$. Observations have found that both the distributions of the SNRs as well as those of the atomic and molecular hydrogen extend approximately upto a radial distance of $R\sim16$ kpc (Case $\&$ Bhattcharya 1996, Gordon $\&$ Burton 1976). Regarding the vertical distributions, detailed studies have found that most of the SNRs and the molecular hydrogen are confined within the region $\sim \pm200$ pc from the Galactic plane (see e.g Stupar et al. 2007, Bronfman et al. 1988). The distribution of atomic hydrogen also follow a similar structure but with a thin long tail extending as far as $\sim 700$ pc from the plane (Dickey $\&$ Lockman 1990). Therefore, in our calculations we consider an infinitely thin disk approximation for both the background sources and the matter distributions. Such an approximation can also be found in some earlier works like Webber et al. 1992, Seo $\&$ Ptuskin 1994 etc.

For the nearby discrete sources, we consider only known SNRs located within a distance of $1.5$ kpc from the Earth. They are listed in Table 1 along with their estimated distances and ages. They are the sources which are expected to produce significant temporal fluctuations in the CR densities at the Earth (Thoudam 2006). The diffusion region for CRs from these sources is assumed to be of infinite dimensions. This assumption is based on the fact that CRs from nearby sources do not effectively see the presence of any Galactic boundaries (both in the radial as well as vertical directions) due to their much smaller propagation times to the Earth compared to their escape timescales from the Galaxy boundaries (Thoudam 2007b).

One more assumption that we make in our model is that since the Earth is reported to be only $\sim 15$ pc away from the Galactic median plane (Cohen 1995), we will simply consider that the Earth is located on the median plane itself in all our calculations.
\begin{table}
\centering
\caption{Parameters of known SNRs located within a distance of $1.5$ kpc from the Earth (See the references given in Thoudam 2007a).}
\begin{tabular}{@{}lllrrlrlr@{}}
\hline
SNR &   Distance$(kpc)$      &Age$(yr)$\\         
\hline
G65.3+5.7   	&     1.0  &   14000\\
G73.9+0.9   		&     1.3  &   10000\\
Cygnus Loop 		&     0.4  &   14000\\
HB21        		&     0.8  &   19000\\
G114.3+0.3  		&     0.7  &   41000\\
CTA1        		&     1.4  &   24500\\
HB9         		&     1.0  &   7700 \\
S147				&	   0.8 &   4600\\
Vela        		&     0.3  &   11000 \\
G299.2-2.9  	&     0.5  &   5000\\
SN185				&	   0.95&	 1800\\
Monogem     		&     0.3  &   86000\\
Geminga     	&     0.15  &   340000\\
\hline
\end{tabular}
\end{table}

\section {CRs from a discrete point source}
\subsection{CR primaries}
In the diffusion model, neglecting convection, energy losses and particle reacceleration processes, the propagation of CR primaries in the Galaxy can be represented by the equation
\begin{equation}
\nabla\cdot(D_p\nabla N_p)-2hnv_p\sigma_p\delta(z)N_p+\mathbb{Q}_p=\frac{\partial N_p}{\partial t}
\end{equation} 
where the subscript $p$ denotes the primary nuclei, $N_p(\textbf{r},E,t)$ is the differential number density at a distance $\textbf{r}$ at time $t$, $E$ is the kinetic energy per nucleon of the nuclei, $D_p(E)$ is the diffusion coefficient and $n$ is the target density both of which are assumed to be constant in the Galaxy, $\sigma_p$ is the primary spallation crossection assumed to be independent of energy, $v_p$ is the primary velocity and $\mathbb{Q}_p(\textbf{r},E,t)=Q_p(E)\delta(\textbf{r})\delta(t-t_0)$ is the particle production rate from the source. In Eq. (1), we neglect the yield of the primaries from the fragmentation of heavier nuclei.

We are also interested in the study of the secondary nuclei which are produced by the nuclear interactions of the primaries with the interstellar medium (ISM). Since the energy per nucleon is an almost conserved quantity in the fragmentation process, we will be dealing with the energy per nucleon in all our relations rather than the total kinetic energy of the nuclei. Using Green's function technique and performing proper fourier and laplace transforms, the exact unbounded solution of Eq. (1) for a source having particle release time $t_0$ is obtained as (see Appendix A)
\begin{align}
N_p(\textbf{r},E,t)=\frac{Q_p(E)e^{-\left[\frac{r^2}{4D_p(t-t_0)}\right]}}{8\pi D_p^{3/2}(t-t_0)}\Biggl\{\frac{e^{- q^2/\left[4(t-t_0)\right]}}{\sqrt{\pi (t-t_0}}-be^{bq+b^2(t-t_0)}\mathrm{erfc}\left(b\sqrt{t-t_0}+\frac{q}{2\sqrt{t-t_0}}\right)\Biggl\}
\end{align}
\\
where $q=|z|/\sqrt{D_p}$ , $b=2hnc\sigma_p/2\sqrt{D_p}$ and we have taken $v_p\approx c$, the velocity of light since we are dealing with high energy particles. The particle flux can then be calculated using $I_p(\textbf{r},E,t)\approx (c/4\pi)N_p(\textbf{r},E,t)$. It should be noted that we assume the source spectrum to be $Q_p(E)=A_pq_p(T)$ with $q_p(T)$ given by
\begin{equation}
q_p(T)=k(T^2+2Tm_p)^{-(\Gamma+1)/2}(T+m_p)
\end{equation}
where $T=A_pE$ represents the total kinetic energy of the nuclei, $A_p$ is the mass number, $m_p$ is the mass energy, $\Gamma$ is the spectral index and $k$ is the normalization constant. 

\subsection{CR secondaries}
The transport of CR secondaries in the Galaxy also follow an equation similar to that of the primaries given above as
\begin{equation}
\nabla\cdot(D_s\nabla N_s)-2hnv_s\sigma_s\delta(z)N_s+\mathbb{Q}_s=\frac{\partial N_s}{\partial t}
\end{equation} 
where the subscript $s$ represents the secondary nuclei and all the quantities have the similar definitions as in Eq. (1). In our model, we assume that the secondaries are the results of fragmentation of one or more heavier primaries and we neglect the production of secondaries at the source. For secondaries originating from a single type of primary, we can write the source term in Eq. (4) as 
\begin{equation}
\mathbb{Q}_s(\textbf{r},E,t)=2hnc\delta(z)\int^{\infty}_{E}\frac{d}{dE^\prime}\sigma_{ps}(E,E^\prime) N_p(\textbf{r},E^\prime,t)dE^\prime
\end{equation}
We can approximate the differential production crossection $d\sigma_{ps}(E,E^\prime)/dE^\prime$ of an $s$-type nuclei of energy per nucleon $E$ by the fragmentation of a $p$-type nuclei of energy per nucleon $E^\prime$ by a delta function as 
\begin{equation}
\frac{d}{dE^\prime}\sigma_{ps}(E,E^\prime)=\sigma_{ps}\delta(E^\prime-E)
\end{equation}
where $\sigma_{ps}$ denotes the total fragmentation crossection of $p$ to $s$. This simplifies Eq. (4) as
\begin{equation}
\mathbb{Q}_s(\textbf{r},E,t)=2hnc\sigma_{ps}\delta(z)N_p(\textbf{r},E,t)
\end{equation}
Note that here $N_p(\textbf{r},E,t)$ is given by Eq. (2). For a secondary source term given by Eq. (7) we can easily obtain the solution of Eq. (4) at the spatial location $\textbf{r}=0$ as
\begin{eqnarray}
N_s(E,t)=2hnc\sigma_{ps}\int_0^tdt_0\int dV \mathrm{exp}\left[-\frac{(\textbf{r}-\textbf{r}^\prime)^2}{4D_s(t-t_0)}\right]\left[\frac{1}{8\pi^{3/2}\left[D_s(t-t_0)\right]^{3/2}}-\frac{2hnc\sigma_s}{16D_s^2(t-t_0)}\right]\mathrm{exp}\left[\frac{4h^2n^2c^2\sigma_s^2(t-t_0)}{4D_s}\right]\nonumber\\
\times \mathrm{erfc}\left[\frac{2hnc\sigma_s\sqrt{t-t_0}}{2\sqrt{D_s}}\right]N_p(\textbf{r}^\prime,E,t)\delta(z^\prime)
\end{eqnarray} 
where $\textbf{r}^\prime$ denotes the position of the primaries with respect to the point source which in cylindrical coordinates yield an integral over the volume element as
\begin{equation}
\int dV=\int_0^{\infty}r^\prime dr^\prime\int_0^{2\pi}d\phi^\prime \int_{-\infty}^{\infty} dz^\prime
\end{equation}
Eq. (8) gives the CR secondary density at $\textbf{r}=0$, which is considered here as the position of the observer, due to a single type of primary emitted from a point source located at a distance $\textbf{r}$ from the observer. If a particular secondary species is produced as the result of fragmentations of more than one type of heavier primaries, the total number of secondaries produced is obtained by simply adding the contributions from the different primary species as
\begin{equation}
N_s^{Tot}=\displaystyle\sum_{j} N_s^j
\end{equation}
where $j$ denotes the primary species.  

\section {CRs generated by the background sources}
\subsection{CR primaries}
CRs from the background sources are assumed to follow a steady state propagation equation in the Galaxy as given below
\begin{equation}
\nabla\cdot(D_p\nabla N_p)-2hnv_p\sigma_p\delta(z)N_p=-S_p
\end{equation}
where $S_p(\textbf{r},E)$ represents the source term. For the presence of a Galactic vertical boundary at $z=\pm H$, the solution of Eq. (11) for a uniform source distribution $S_p(\textbf{r},E)=\Re Q_p(E)\delta(z)$ extended upto a radial distance $R$ in the Galactic plane is given by (see Appendix B)
\begin{equation}
N_p(z,E)=\frac{R\Re Q_p(E)}{2D_p}\int^\infty_0\frac{\mathrm{sinh}[K(H-z)]}{\mathrm{sinh}(KH)\left[K\mathrm{coth}(KH)+\frac{2hnc\sigma_p}{2D_p}\right]}\times \mathrm{J_1}(KR)dK
\end{equation}
where $\mathrm{J_1}$ is the Bessel function of order 1 and $\Re=25$ Myr$^{-1}$ kpc$^{-2}$ denotes the supernova explosion rate in the Galaxy (Grenier 2000). As already mentioned in section 2, a major fraction of the CRs reaching the Earth are liberated by sources located within a distance which is of the order of the vertical height $H=\pm 5$ kpc (see Taillet $\&$ Maurin 2003). Therefore, considering the fact that the Earth is located at a distance of $\sim 8.5$ kpc away from the center of the Galaxy and that the sources are distributed upto a radial distance of $R\sim 16$ kpc, Eq. (12) can be used to obtain the CR density at the Earth by setting $z=0$. 

\subsection{CR secondaries}
The density of CR secondaries which are produced by the fragmentation of the heavier primaries originated from a stationary source distribution can be obtained similar to that of the primaries as
\begin{equation}
N_s(z,E)=2hnc\sigma_{ps}N_p(z,E)\frac{R}{2D_s}\int^\infty_0\frac{\mathrm{sinh}[K(H-z)]}{\mathrm{sinh}(KH)\left[K\mathrm{coth}(KH)+\frac{2hnc\sigma_s}{2D_s}\right]}\times \mathrm{J_1}(KR)dK
\end{equation} 
where we have taken the secondary source term as $S_s(\textbf{r},E)=2hnc\sigma_{ps}\delta(z)N_p(\textbf{r},E)$. Taking $z=0$, Eq. (13) gives the secondary densities at the Earth.
Eq. (13) shows that for very large $D_s$, the secondary to primary ratio for a stationary source distribution follows
\begin{equation}
\frac{N_s}{N_p}\propto \frac{1}{D_s}
\end{equation}
The above relation shows that if CRs are liberated by stationary sources which are distributed uniformly, the s$/$p can be used to estimate the CR diffusion cofficient in the Galaxy. However, this approach may fail in cases where the influence of nearby discrete sources is significant. It is because the presence of strong nearby sources can significantly affect the CR fluxes (mainly the primaries) which can subsequently affect the s$/$p ratio (see B$\mathrm{\ddot{u}}$sching et al. 2005). How far the measured ratio will deviate from Eq. (14) actually depends on the ages and distances of the nearby sources. In the next section, we will investigate the variations that one can expect in the CR densities due to the presence of a single nearby source.

\section{CR density variations due to a single nearby source}
\begin{figure}
\centering
\includegraphics*[width=0.31\textwidth,angle=270,clip]{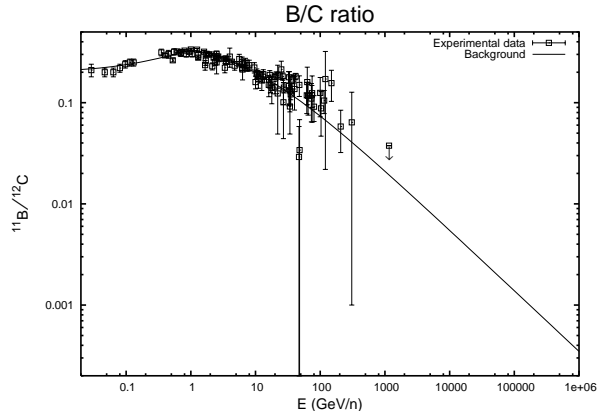}
\caption{\label {fig1} B/C ratio at the Earth due to the background CRs (solid line). Model parameters: Oxygen$/$carbon source abundance ratio O$/$C $=1.4$ and $\Phi=500$ MV. Experimental data are taken from Swordy et al. 1990, Panov et al. 2007 and the compilations of different experiments given in Stephens $\&$ Streitmatter 1998.}
\end{figure}
One very important parameter in the study of CR propagation in the Galaxy is the CR diffusion coefficient $D$. The diffusion of CRs is generally considered to be due to scattering either by magnetic field iregularities or by self excited Alfven and hydromagnetic waves. Because of the possible spatial variations in the scattering processes, the value of $D$ may be different at different locations in the Galaxy. However, in the present work we make a simple approximation that $D$ remains constant throughout the Galaxy. As already discussed in the last section, the diffusion coefficient can be determined using the s$/$p ratio sunder the steady state model. Though the value thus obtained may not represent the true value because of the presence of nearby discrete sources, to begin with, we assume that
\begin{eqnarray}
D_i(E_i)=D_0\left(\frac{E_0}{E_i}\right)^{0.6}\quad;\quad (E_i<E_0) \nonumber\\
\qquad =D_0\left(\frac{E_i}{E_0}\right)^{0.6}\quad;\quad (E_i>E_0)
\end{eqnarray}
where $E_i$ denotes the kinetic energy per nucleon of a particular nuclear species denoted by the subscript $i$ and the values of $D_0$ and $E_0$ are chosen so that Eq. (14) fits the observed boron$/$carbon (B/C) ratio at $1$ GeV/n (see Figure 1). We obtain $D_0=2.9\times 10^{28}$ cm$^2$s$^{-1}$ and the particle rigidity (corresponding to energy $E_0$) with charge $Z_i$ and mass number $A_i$ as $\rho_0=A_iE_0/Z_i=3$ GV. We assume that the boron secondaries ($^{11}$B) are produced by the nuclear interactions of the $^{12}$C and $^{16}$O progenitors with the ISM. The values for the nuclear fragmentation crossections ($\sigma_p,\sigma_s,\sigma_{ps}$) used in the present work are taken from Webber et al. 1992 $\&$ 1998 and are listed in Table 2. We assume the ISM target density which consists mainly of hydrogen atom to be $n=1$ cm$^{-3}$ and we take into account the solar modulation effect by taking the modulation parameter to be $\Phi=500$ MV. Note that the necessity for the break in the diffusion coefficient at $E_0$ to reproduce the peak in the observed data somewhere around $1$ GeV/n is consistent with the earlier studies based on the diffusive and leaky box propagation models. It is also worth mentioning that in the case of reacceleration models, the peak can be explained using a single power-law diffusion coefficient without assuming any break in energy (see e.g., Seo $\&$ Ptuskin 1994). However, the observed decrease of the secondary abundances with energy above about $1$ GeV/n suggests that reacceleration in the ISM cannot be regarded as the dominant process for particles with energies $\gtrsim 1$ GeV/n (see e.g., Hayakawa 1969). Therefore, by neglecting particle reacceleration as well as other possible low energy effects like convection and energy losses, we assume that Eq. (1) properly describes the propagation of CRs in the Galaxy at least for energies greater than $\sim 1$ GeV/n. 

\begin{figure}
\centering
\includegraphics*[width=0.3\textwidth,angle=270,clip]{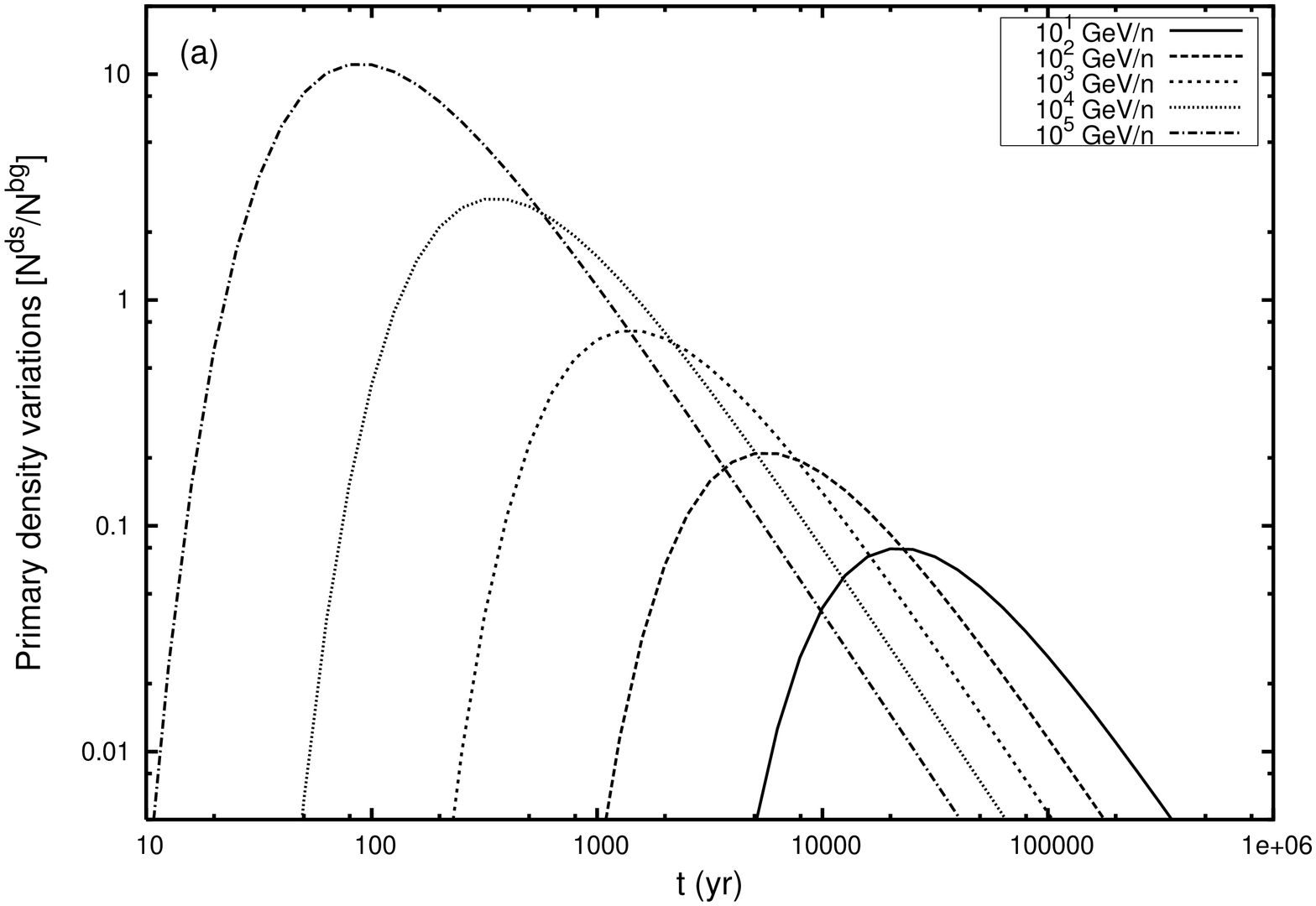}
\includegraphics*[width=0.3\textwidth,angle=270,clip]{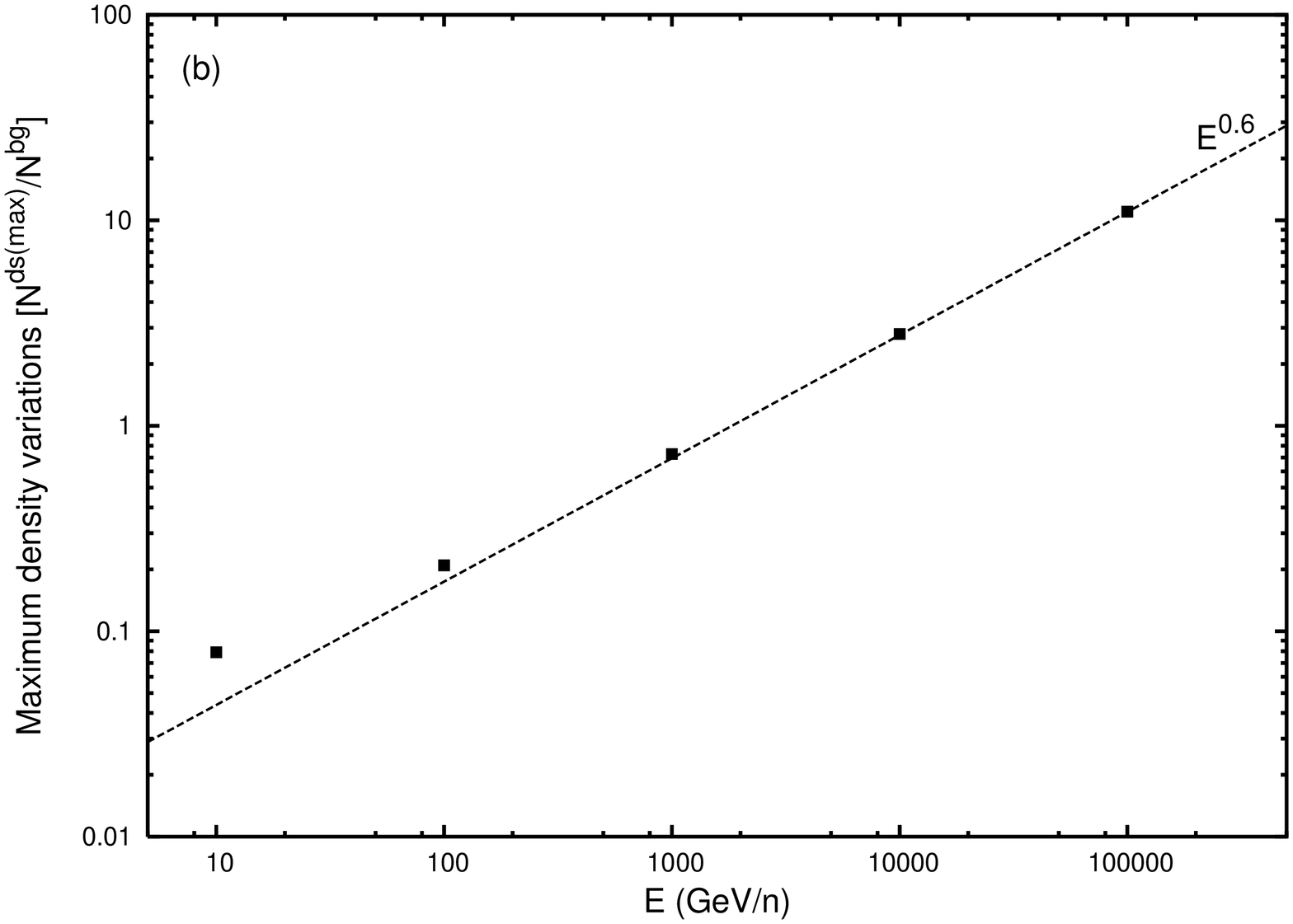}
\caption{\label {fig1} Left: $^{12}$C temporal density variations expected at the Earth for energies $(10-10^5)$ GeV/n due to the presence of a nearby discrete source. For the calculation, we assume the source distance $r=0.2$ kpc, $t_0=0$, $\Gamma=2.25$, $\Re=25$ Myr$^{-1}$ kpc$^{-2}$ and $\Phi=500$ MV. The variations are calculated from the ratio $N^{ds}/N^{bg}$, where $N^{ds}$ and $N^{bg}$ denote the CR densities due to the discrete and the background sources respectively. Right: Maximum density variations (solid squares) expected at different energies, i.e. the peak values for the curves shown in the left panel. The dashed straight line in the figure corresponds to $E^{0.6}$ which shows that the maximum density variation increases with energy with a slope close to the diffusion coefficient index. $N^{ds(max)}$ denotes the maximum density the source can produce at the Earth (see text for details).} 
\end{figure}

Figure 2 (left) shows the expected variations of $^{12}$C densities at the Earth due to the presence of a nearby discrete source. We choose the distance to the source as $0.2$ kpc and  the source spectral index as $\Gamma=2.25$. The density variations are obtained from the ratio $N^{ds}/N^{bg}$ where $N^{ds}$ and $N^{bg}$ denote the densities due to the discrete (given by Eq. 2) and the background (Eq. 12) sources respectively. In Figure 2 (left), the curves from right to left represent for energies $(10-10^5)$ GeV/n. We can see that higher energy particles can produce variations of much larger amplitudes than the lower energy ones. This can be understood from Figure 2 (right) where we have plotted only the maximum density variations (represented by the solid squares) at different energies, i.e. the peak values for each of the curves shown in Figure 2 (left). Neglecting particle losses due to nuclear fragmentations, the maximum density of particles of energy $E$ due to a point source located at a distance $r$ can be found at an age $t_{max}(E)=r^2/6D_p(E)$ and is given by $N^{ds(max)}(E)\propto Q_p(E)$. The background CRs as given by Eq. (12) follows $N^{bg}(E)\propto Q_p(E)/D_p(E)$. Therefore, the maximum deviation that a discrete source can produce can be obtained as
\begin{equation}
\frac{N^{ds(max)}}{N^{bg}}\propto D_p(E)
\end{equation}
For reference, we have also drawn a straight line $(\propto E^{0.6})$ in Figure 2 (right) which shows that the maximum density variation increases with energy  with a slope equal to the index of the CR diffusion coefficient. The deviation from the straight line at lower energies is due to the increasing importance of the nuclear fragmentations compared to the diffusion processes at these energies. Therefore, if a nearby source is present in the local ISM, we should expect larger density variations for particles which diffuse faster in the Galaxy. However in actual practice, since the observations are made at a particular time and location, the actual variation at an energy $E$ depends on the age(s) and distance(s) of the nearby source(s).   
 
\section{Application to the nearby SNRs}
\begin{figure}
\centering
\includegraphics*[width=0.31\textwidth,angle=270,clip]{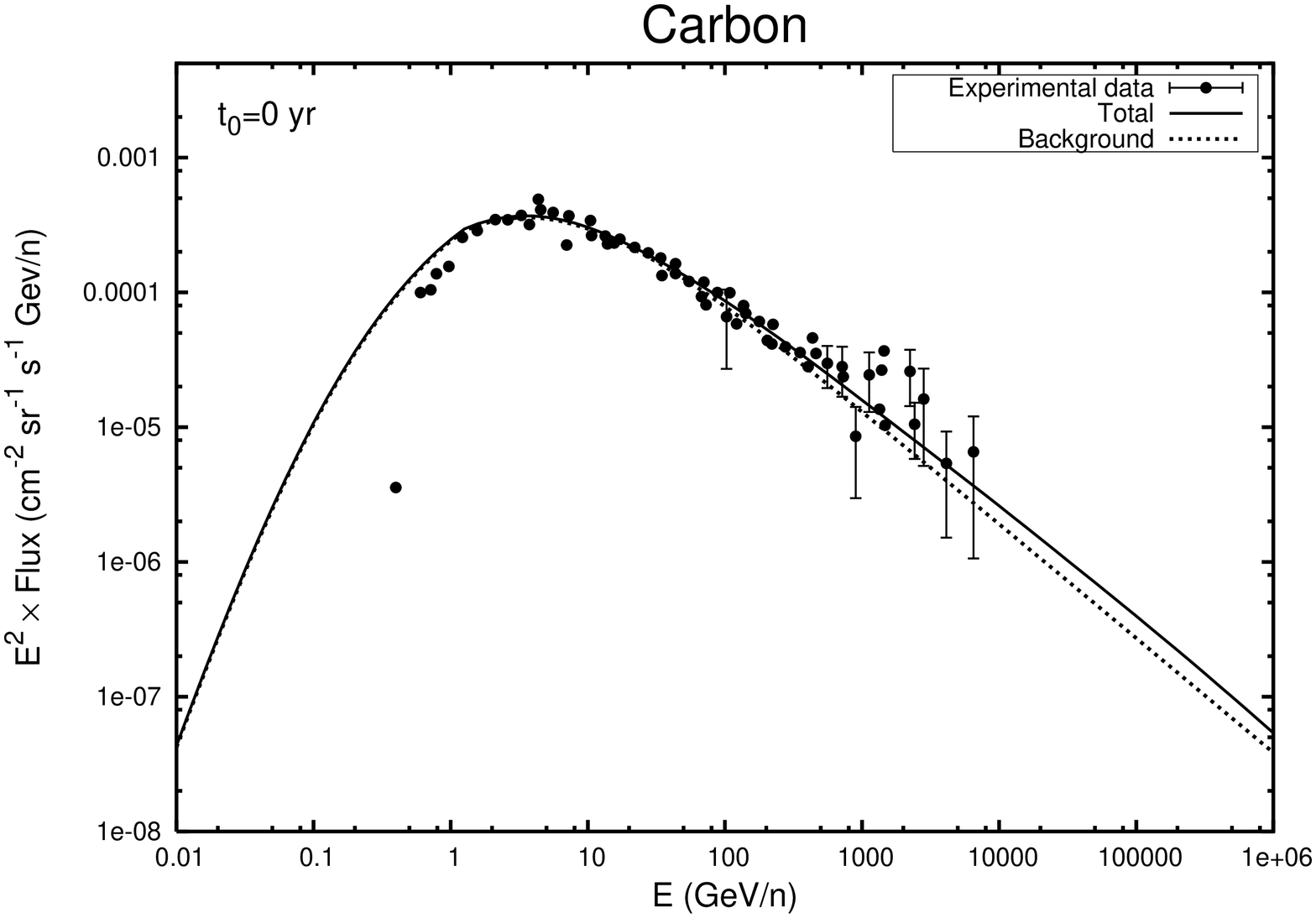}
\includegraphics*[width=0.31\textwidth,angle=270,clip]{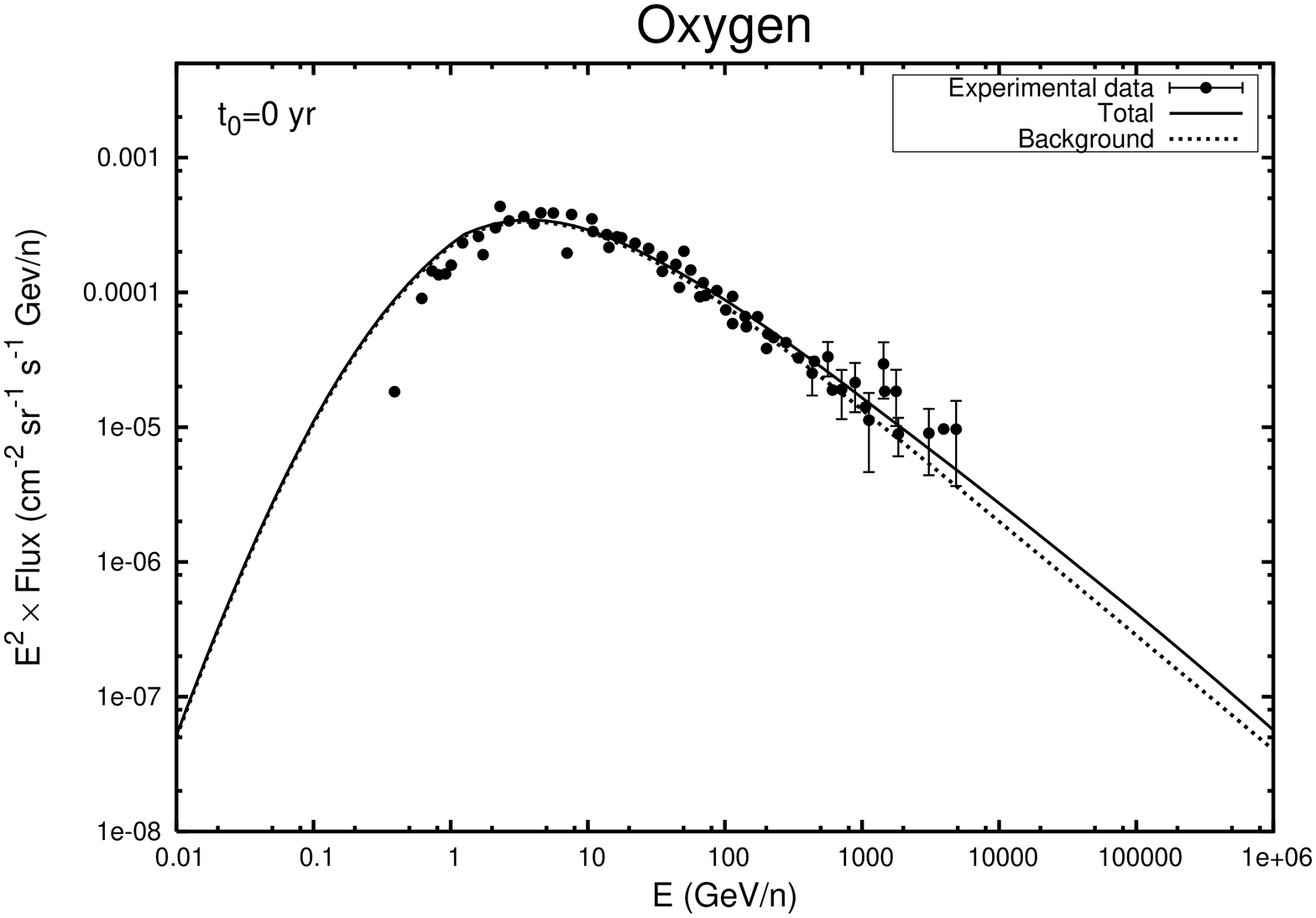}
\caption{\label {fig1} Left: Calculated $^{12}$C spectra normalized to the observed data at $10$ GeV/n. Right: $^{16}$O spectra calculated for the O/C source abundance ratio of 1.4. Other model parameters: $t_0=0$, $\Gamma=2.25$, $\Re=25$ Myr$^{-1}$ kpc$^{-2}$, $\Phi=500$ MV. The dotted lines represent the background CRs and the solid lines the total flux which also include the contributions of the nearby SNRs listed in Table 1. Data points are taken from the results of different experiments given in Zei et al. 2007.}
\end{figure}
\begin{figure}
\centering
\includegraphics*[width=0.31\textwidth,angle=270,clip]{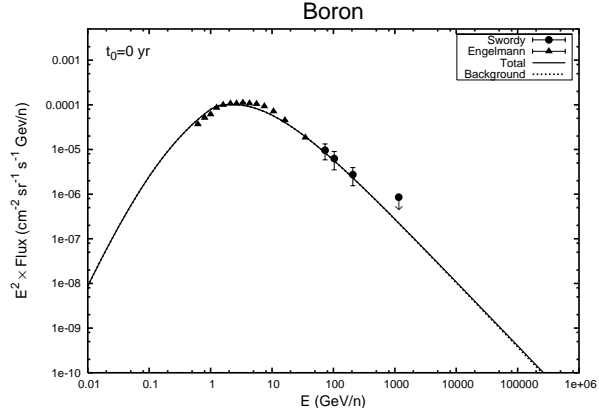}
\caption{\label {fig2}$^{11}$B secondary spectra at the Earth calculated using the $^{12}$C and $^{16}$O spectra shown in Figure 3. The background flux (dotted line) almost overlap with the total flux (solid line) since the total contribution of the nearby SNRs is almost neglible. Solar modulation parameter $\Phi=500$ MV. Data points: Swordy et al. 1990 and Engelmann et al. 1990.}
\end{figure}
\begin{figure}
\centering
\includegraphics*[width=0.31\textwidth,angle=270,clip]{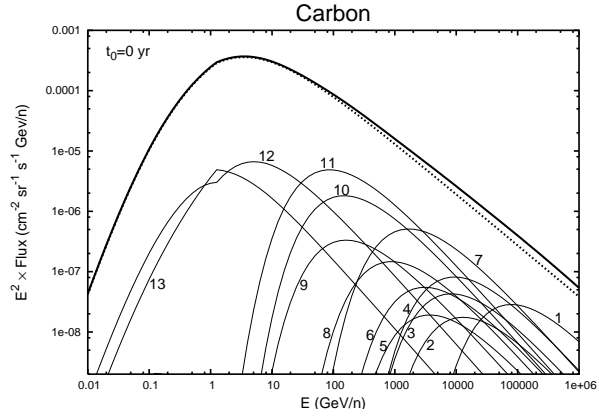}
\caption{\label {fig2} Same as the $^{12}$C spectra shown in Figure 3 (left), but now showing the contributions from the individual SNRs. Dotted line: Background spectra. Thick solid line: Total spectra. The thin solid lines labelled as $(1-13)$ represent the contributions from the individual SNRs listed in Table 1: $1-$ SN185, $2-$ G73.9+0.9, $3-$ HB9, $4-$ S147, $5-$ CTA1, $6-$ G65.3+5.7, $7-$ G299.2-2.9, $8-$ HB21, $9-$ G114.3+0.3, $10-$ Cygnus Loop, $11-$ Vela, $12-$ Monogem and $13-$ Geminga.}
\end{figure}

\begin{table}
\centering
\caption{Fragmentation crossections used in our calculations (taken from Webber et al. 1992 $\&$ 1998). The subscripts $C,O$ $\&$ $B$ denote the $^{12}$C, $^{16}$O $\&$ $^{11}$B nuclei respectively and $CB$ $\&$ $OB$ represent the $^{12}$C$\rightarrow ^{11}$B $\&$ $^{16}$O$\rightarrow ^{11}$B processes respectively.}
\begin{tabular}{@{}lllrrlrlr@{}}
\hline
Crossection & Value (mbarn)\\
\hline
$\sigma_C$  &   250\\
$\sigma_O$  &   308\\
$\sigma_B$  &   232\\
$\sigma_{CB}$	&76.8\\
$\sigma_{OB}$	&38.5\\
\hline
\end{tabular}
\end{table}
In this section, we will investigate the possible effects of the presence of nearby SNRs on the observed primary and secondary spectra and also, the subsequent effects on the s/p ratios. For the s/p ratio study, we choose the B/C ratio as an example since it is found to be the most well-measured ratio among all the available s/p ratios.

In our study, we consider only the $^{12}$C and $^{16}$O primaries since boron secondaries are found to be predominantly produced by the nuclear interactions of these two species with the ISM. They contribute roughly $50\%$ and $25\%$ respectively of the overall boron produced (Webber et al. 1998). The calculated spectra (total as well as background) for the CR primaries and secondaries are shown in Figures $3\&4$ respectively along with the experimental data. The total spectra represented by the solid lines in the figures correspond to the total CR densities which are given by
\begin{equation}
N_C=N_C^{bg}+\displaystyle\sum_{i} N_C^{ds_i}
\end{equation}
\begin{equation}
N_O=N_O^{bg}+\displaystyle\sum_{i} N_O^{ds_i}
\end{equation} 
\begin{equation}
N_B=N_{CB}^{bg}+N_{OB}^{bg}+\displaystyle\sum_{i} N_{CB}^{ds_i}+\displaystyle\sum_{i} N_{OB}^{ds_i}
\end{equation}
where the subscripts $C,O$ and $B$ denote the $^{12}$C, $^{16}$O and $^{11}$B nuclei respectively, the superscripts $bg$ and $ds$ denote the background (represented by the dotted lines in the figures) and the discrete components respectively, and the summations are over the discrete sources $i$ listed in Table 1. The subscripts $CB$ and $OB$ represent the $^{12}$C$\rightarrow ^{11}$B and $^{16}$O$\rightarrow ^{11}$B processes respectively. We adopt a particle release time of $t_0=0$ for the discrete sources. The source spectral index is taken as $\Gamma=2.25$ and the value of the source normalization constant $k$ is chosen such that the resulting total $^{12}$C spectrum is normalized to the observed spectrum at $10$ GeV/n for the supernova explosion rate of $\Re=25$ Myr$^{-1}$ kpc$^{-2}$ in the Galaxy. This is shown in Figure 3 (left). The $^{16}$O spectra calculated using the source abundance ratio of O/C $=1.4$ is shown in Figure 3 (right). Note that Engelmann et al. 1990 had used a source abundance ratio of $1.24$ to reproduce the observed C and O abundances using the simple leaky box CR propagation model. The experimental data in Figure 3 are taken from the compilation of different experiments given in Zei et al. 2007. The secondary boron spectrum, calculated using the $^{12}$C and $^{16}$O primary spectra shown in Figure 3, is found to explain the observed boron data quite well. This is shown in Figure 4 where the experimental data are taken from Swordy et al. 1990 and Engelmann et al. 1990.  

For the primaries, one can notice from Figure $3$ that the inclusion of the nearby SNRs in the study produces noticeable deviations from the background flux for energies above $\sim 100$ GeV/n. For instance, the deviation in the case of $^{12}$C reaches a value of $\sim 45\%$ at energies around $10^5$ GeV/n. But, from Figure $4$ we can see that the deviation in the case of $^{11}$B which is considered here as a purely secondary particle is almost negligible. 
The results for the secondaries obtained in this paper agree quite well with the earlier findings of B$\mathrm{\ddot{u}}$sching et al. 2005, but some major differences can be seen in the case of the primaries. They found a typical primary amplitude variations of $\sim20\%$ at almost all the energies whereas we find almost no variations upto $\sim 100$ GeV/n and beyond that we see a slow increase reaching a value of $\sim 45\%$ at $\sim 10^5$ GeV/n. This can be clearly understood if one examines Figure $5$ in detail where we have plotted the individual components of the overall $^{12}$C spectrum previously shown in Figure $3$ (left). The contributions from the individual SNRs are labelled as $1-13$ (see the figure caption for details). It can be seen that among the SNRs, only the Monogem and the Geminga give the highest contribution below $\sim 100$ GeV/n while the rest of the SNRs contribute mostly above $\gtrsim 100$ GeV/n. Since the CR flux from a point source depends strongly on the age and distance of the source, the low energy CRs from the nearby SNRs have not yet reached us effectively except those coming from the Monogem and Geminga SNRs. But still, the maximum contributions of Monogem and Geminga below $\sim 100$ GeV/n are found to be approximately two orders of magnitude less than the overall flux. Above $\gtrsim 100$ GeV/n, the major contributors are Vela, G$299.2$-$2.9$ and SN$185$, each one of them contributing around $10\%$ of the total flux at different energy regions. Adding the contributions of the other SNRs also, the total contribution from the nearby known SNRs comes to $\sim 31\%$ of the overall CR flux which corresponds to $\sim 45\%$ deviation from the background level at energies $\sim (10^4-10^6)$ GeV/n.
\begin{figure}
\centering
\includegraphics*[width=0.375\textwidth,angle=270,clip]{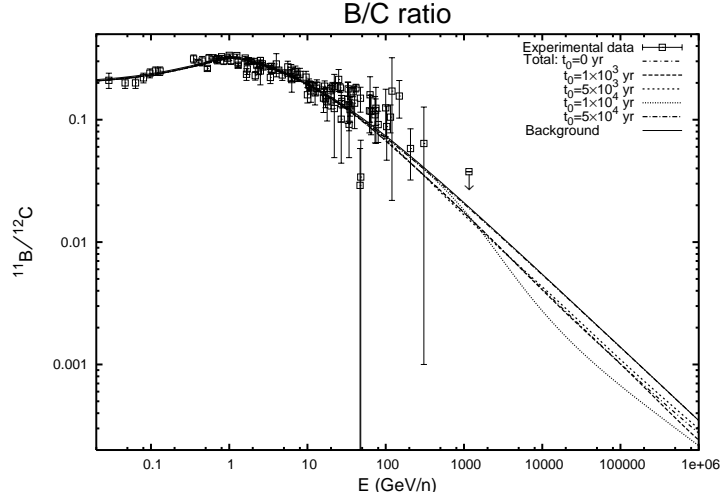}
\caption{\label {fig3} Expected B/C ratios at the Earth in the presence of the nearby SNRs listed in Table 1, calculated for different particle injection times $t_0=(0-5\times 10^4)$ yr as indicated in the key box. Experimental data are the same as in Figure 1. For reference, we have also shown the B/C ratio arising purely from the CR background (solid line). Model parameters: O/C $=1.4$, $\Gamma=2.25$, $\Re=25$ Myr$^{-1}$ kpc$^{-2}$ and $\Phi=500$ MV.}
\end{figure}
\begin{figure}
\centering
\includegraphics*[width=0.31\textwidth,angle=270,clip]{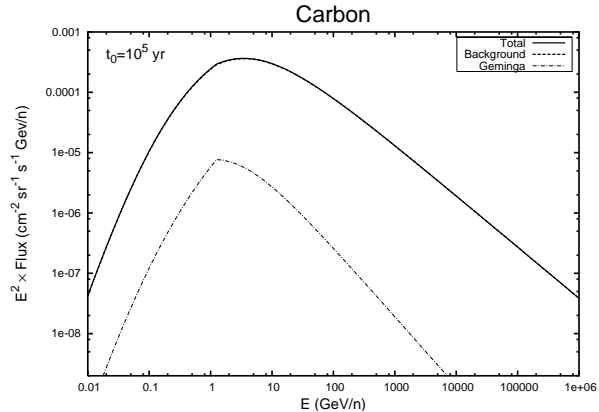}
\caption{\label {fig2}Same as Figure 5, but calculated for $t_0=10^5$ yr. Among all the SNRs listed in Table 1, only Geminga with an estimated age of $\sim 3.4\times 10^5$ yr have released CRs in the ISM. Dot-dashed line: Geminga. Solid line: Total flux. Thick-dashed line: Background flux (not clearly visible because it almost overlap with the total flux).}
\end{figure}

The deviation in the $^{12}$C primary spectra is expected to give a direct implication on the total B/C ratio. This is shown in Figure $6$ where we have plotted the ratios ($N_B/N_C$) expected in the presence of the nearby SNRs. For reference, we have also plotted the ratio expected purely from the CR background (solid line). The experimental data are taken from Swordy et al. 1990, Panov et al. 2007 and the compilations of different experiments given in Stephens $\&$ Streitmatter 1998. Strictly speaking, it is not the age $t$ alone which determines the contribution of a discrete source but the propagation time $\Delta_t=t-t_0$ of the particles after they are released from the source. But, the value of $t_0$ is not exactly known. It may be even different for different sources though we assume the same value for all the sources in the present work. So, in Figure $6$ we have considered various particle release times i.e. $t_0=(0, 10^3, 5\times 10^3, 10^4$ $\&$ $5\times 10^4)$ yr. Below $\sim 100$ GeV, it can be seen that at all the $t_0'$s the ratios show almost zero deviations from the background ratio. Any deviation, if exists, are seen at energies $\gtrsim 100$ GeV. At $t_0<10^4$ yr, we see deviations of magnitude $\sim(16-26)\%$ in the energy range of $\sim (10^3-10^6)$ GeV/n. At $t_0=10^4$ yr, the ratio shows the maximum deviation reaching a value of $\sim 52\%$ at $\sim 6\times 10^4$ GeV/n and for $t_0>10^4$ yr, say at $t_0=5\times 10^4$ yr, the effect of the nearby SNRs becomes negligible producing almost zero deviation from the background ratio. It is worth mentioning at this point that detailed studies based on the diffusive shock acceleration in SNRs suggest that the highest energy particles start leaving the source region already at the beginning of the Sedov phase (Berezhko et al. 1996), but the major fraction of the accelerated CRs remain confined for almost around $10^5$ yr for a typical interstellar hydrogen atom density of $n=1$ cm$^{-3}$. We will come to this point later again in the next section while discussing about the implications of the results obtained in this paper. Just for the sake of completeness, we have plotted the $^{12}$C spectrum for $t_0=10^5$ yr in Figure 7. Here, except Geminga (being an old SNR with age $\sim 3.4\times 10^5$ yr), all the other nearby SNRs have not yet released CRs into the local ISM, thereby, leading to a negligible effect at the Earth.       

\section{Discussions and conclusions}
We have studied the effect of the presence of nearby SNRs on the CR primary and secondary spectra at the Earth. We see strong variations in the primary spectra and almost no variation in the case of the secondaries. The results for the primaries obtained here are quite different from those obtained from the simulation studies given in Erlykin $\&$ Wolfendale (2001) and B$\mathrm{\ddot{u}}$sching et al. (2005). We see variations mostly above $\sim 100$ GeV/n whereas they showed significant fluctuations at all the energies. In fact, their results represent the density fluctuations that one can expect at any arbitrary location in the Galaxy due to the random nature of supernova explosions both in space as well as time. But, as far as the position of the Earth is concerned, the actual variations can be determined only when one incorporates the nearby sources in the study. We, therefore, include the nearby known SNRs in our analysis and found that the primary variations obtained at the Earth show considerable differences from those predicted using the Monte-Carlo simulations. 

Below $\sim 100$ GeV/n, we have found that the effect of the nearby SNRs on the B/C ratio is negligible (see Figure 6). This implies that we can safely rely on the observed ratio to determine the CR diffusion coefficient in the Galaxy particularly below $\sim 100$ GeV/n. Above this energy the observed data will not give a reliable information about the propagation parameter because of the significant contaminations of the background CRs by those coming from the nearby SNRs. One should note that the primary CRs observed at the Earth are liberated from sources located within a short distance which is of the order of the vertical halo height $H$ (Taillet $\&$ Maurin 2003). This can also be understood from Fig. 8 where we have plotted the fraction of $^{12}$C primaries at the Earth originated within a radial distance $r$ for energies $(10-10^4)$ GeV/n. The calculations are performed for $H=5$ kpc. We can see from the figure that for energies less than $100$ GeV/n, $50\%$ ($70\%$) of the total CRs are emitted within a distance of $\sim 3$ kpc ($5$ kpc). This shows that even though the B/C ratio below $\sim 100$ GeV/n can give reliable informations about the diffusion coefficient, the small Galactic region scanned by the CRs reaching the Earth allows them to carry information only for a small fraction of the whole Galaxy (see also Taillet $\&$ Maurin 2003).

We have also studied the effect of different particle release times ($t_0=0-10^5$ yr) from the SNRs on the B/C ratio. At $t_0<10^4$ yr, we have found a deviation of $\sim(16-26)\%$ from the background ratio for energies $\sim (10^3-10^6)$ GeV/n and at $t_0=10^4$ yr we see the maximum deviation reaching a value of $\sim 52\%$ at around $6\times 10^4$ GeV/n. For $t_0>10^4$ yr, say at $t_0=5\times 10^4$ yr, the effect of the nearby SNRs becomes almost negligible, thereby, making the observed B/C data a reliable quantity for determining the CR propagation parameters at all energies. This conclusion is, in fact, supported by the predictions of the diffusive shock acceleration theories in SNRs which show that a major fraction of the accelerated CRs remain confined in SNRs for almost around $10^5$ yr (Berezhko et al. 1996). For such a confinement time, we see that except Geminga all the other local SNRs listed in Table 1 look quite young and  might not have liberated CR particles in the local ISM. This can be understood from Figure 7 where we have plotted the $^{12}$C spectra for $t_0=10^5$ yr.  

Therefore, we can conclude that if one accepts the theoretical view of CRs confinement in SNRs for upto $\sim 10^5$ yr, the effect of the nearby SNRs on the observed CRs is expected to be quite negligible. Then, one can expect the observed s/p ratio to give a good estimation of the CR diffusion coefficient at all energies, but only for the small Galactic region traversed by them before reaching the Earth.             

\begin{figure}
\centering
\includegraphics*[width=0.3\textwidth,angle=270,clip]{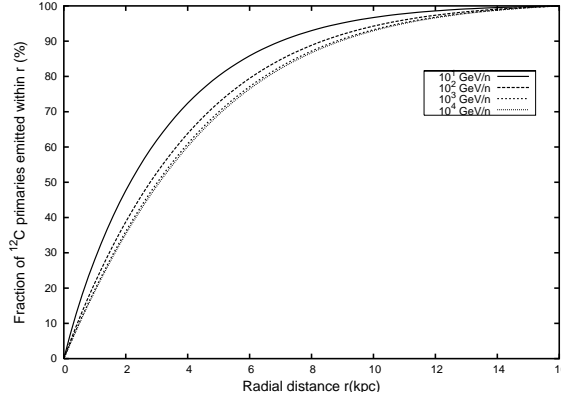}
\caption{\label {fig3} Fraction of $^{12}$C primaries emitted within the radial distance $r$ from the Earth for various energies $(10-10^4)$ GeV/n. We assume the vertical halo height $H=5$ kpc.}
\end{figure}

\appendix
\section{Solution of the time dependent diffusion equation without boundaries}
In rectangular coordinates, the Green's function $G(\textbf{r},\textbf{r}^\prime,t,t^\prime)$ of Eq. (1) satisfies
\begin{equation}
D_p\left(\frac{\partial^2G}{\partial x^2}+\frac{\partial^2G}{\partial y^2}+\frac{\partial^2G}{\partial z^2}\right)-2hnc\sigma_p\delta(z)G-\frac{\partial G}{\partial t}=-\delta(x-x^\prime)\delta(y-y^\prime)\delta(z-z^\prime)\delta(t-t^\prime)
\end{equation} 
Since the paricles are assumed to be liberated at time $t=t^\prime$, Eq. (A1) for $t>t^\prime$ becomes simply
\begin{equation}
D_p\left(\frac{\partial^2G}{\partial x^2}+\frac{\partial^2G}{\partial y^2}+\frac{\partial^2G}{\partial z^2}\right)-2hnc\sigma_p\delta(z)G-\frac{\partial G}{\partial t}=0
\end{equation}
Taking fourier transforms of Eq. (A2) with respect to $x$ and $y$, we obtain
\begin{equation}
-D_pK^2\bar{G}+D_p\frac{\partial^2 \bar{G}}{\partial z^2}-2hnc\sigma_p\delta(z)\bar{G}-\frac{\partial \bar{G}}{\partial t}=0
\end{equation}
where $K^2=k_x^2+k_y^2$ and 
\begin{equation}
\bar{G}(k_x,x^\prime,k_y,y^\prime,z,z^\prime,t,t^\prime)=\int_{-\infty}^\infty dx\int_{-\infty}^\infty dy G(x,x^\prime,y,y^\prime,z,z^\prime,t,t^\prime)e^{ik_xx+ik_yy}
\end{equation}
Now, taking laplace tranform of Eq. (A3) with respect to $t$, we have
\begin{equation}
D_p\frac{\partial^2 \bar{\bar{G}}}{\partial z^2}-\left[D_pK^2+2hnc\sigma_p\delta(z)\right]\bar{\bar{G}}+\delta(z)e^{ik_xx^\prime+ik_yy^\prime}e^{-st^\prime}-s\bar{\bar{G}}=0
\end{equation}
where $\bar{\bar{G}}(k_x,x^\prime,k_y,y^\prime,z,z^\prime,s,t^\prime)=\int_0^\infty \bar{G}e^{-st}dt$ and we have used the condition that at $t=t^\prime$, $G(x,x^\prime,y,y^\prime,z,z^\prime,t,t^\prime)=\delta(x-x^\prime)\delta(y-y^\prime)\delta(z-z^\prime)$. Note that in Eq. (A5), we have set the value of $z^\prime$ equal to $0$ since we will be assumming that the CR sources are located on the Galactic plane itself. Solving Eq. (A5) for the regions above and below the $z=0$ plane by using the proper boundary conditions at $z=\pm\infty$, we get
\begin{equation}
\bar{\bar{G}}(z,s,t^\prime)=\bar{\bar{G}}_0(0,s,t^\prime)e^{-|z|\sqrt{K^2+s/D_p}}
\end{equation}  
The continuity equation at $z=0$ can be obtained by integrating Eq. (A5) over $z$ around $0$ as
\begin{equation}
D_p\left[\frac{\partial \bar{\bar{G}}}{\partial z}\right]^{+0}_{-0}-2hnc\sigma_p\bar{\bar{G}}_0+e^{ik_xx^\prime+ik_yy^\prime}e^{-st^\prime}=0
\end{equation}
Solving for $\bar{\bar{G}}_0$ from Eqs. (A6 $\&$ A7), we get
\begin{equation}
\bar{\bar{G}}_0(0,s,t^\prime)=\frac{e^{ik_xx^\prime+ik_yy^\prime}e^{-st^\prime}}{\left[2D_p\sqrt{K^2+s/D_p}+2hnc\sigma_p\right]}
\end{equation} 
and substituting it back to Eq. (A6), we get 
\begin{equation}
\bar{\bar{G}}(z,s,t^\prime)=\frac{e^{-|z|\sqrt{K^2+s/D_p}}\times e^{ik_xx^\prime+ik_yy^\prime}e^{-st^\prime}}{\left[2D_p\sqrt{K^2+s/D_p}+2hnc\sigma_p\right]}
\end{equation}
Taking an inverse laplace's transform of Eq. (A9), we get (see Abramovitz $\&$ Stegun 1964)
\begin{equation}
\bar{G}(k_x,x^\prime,k_y,y^\prime,z,z^\prime,t,t^\prime)=\frac{e^{ik_xx^\prime+ik_yy^\prime}e^{-D_pK^2(t-t^\prime)}}{2\sqrt{D_p}}\left\lbrace\frac{e^{-q^2/\left[4(t-t^\prime)\right]}}{\sqrt{\pi (t-t^\prime)}}-be^{bq+b^2(t-t^\prime)}\mathrm{erfc}\left(b\sqrt{t-t^\prime}+\frac{q}{2\sqrt{t-t^\prime}}\right)\right\rbrace
\end{equation}
where $q=|z|/\sqrt{D_p}$ and $b=2hnc\sigma_p/2\sqrt{D_p}$. Further taking an inverse fourier transform of Eq. (A10) gives
\begin{eqnarray}
G(x,x^\prime,y,y^\prime,z,z^\prime,t,t^\prime)=\frac{e^{-\left[\frac{(x^\prime-x)^2+(y^\prime-y)^2}{4D_p(t-t^\prime)}\right]}}{8\pi D_p^{3/2}(t-t^\prime)}\left\lbrace\frac{e^{- q^2/\left[4(t-t^\prime)\right]}}{\sqrt{\pi (t-t^\prime)}}-be^{bq+b^2(t-t^\prime)}\mathrm{erfc}\left(b\sqrt{t-t^\prime}+\frac{q}{2\sqrt{t-t^\prime}}\right)\right\rbrace
\end{eqnarray}
Using the Green's function given by Eq. (A11), we can easily obtain the CR density due to an arbitrary source $\mathbb{Q}_p(\textbf{r},E,t)$ as
\begin{equation}
N_p(\textbf{r},E,t)=\int^{\infty}_{-\infty}d\textbf{r}^\prime\int^t_{-\infty}dt^\prime G(\textbf{r},\textbf{r}^\prime,t,t^\prime)\mathbb{Q}_p(\textbf{r}^\prime,,t^\prime)
\end{equation}
For a point source of the form $\mathbb{Q}_p(\textbf{r}^\prime,E,t^\prime)=Q_p(E)\delta(\textbf{r}^\prime)\delta(t^\prime-t_0)$, we obtain a solution given by 
\begin{align}
N_p(\textbf{r},E,t)=\frac{Q_p(E)e^{-\left[\frac{x^2+y^2}{4D_p(t-t_0)}\right]}}{8\pi D_p^{3/2}(t-t_0)}\Biggl\{\frac{e^{- q^2/\left[4(t-t_0)\right]}}{\sqrt{\pi (t-t_0}}-be^{bq+b^2(t-t_0)}\mathrm{erfc}\left(b\sqrt{t-t_0}+\frac{q}{2\sqrt{t-t_0}}\right)\Biggl\}
\end{align}
where in cylindrical coordinates, we can just write $x^2+y^2=r^2$.

\section{ Solution of the steady state diffusion equation with vertical Galactic boundaries}
The Green's function of the steady state diffusion equation [Eq. (10)] in rectangular coordinates satisfies
\begin{equation}
D_p\left(\frac{\partial^2G}{\partial x^2}+\frac{\partial^2G}{\partial y^2}+\frac{\partial^2G}{\partial z^2}\right)-2hnc\sigma_p\delta(z)G=-\delta(x-x^\prime)\delta(y-y^\prime)\delta(z-z^\prime)
\end{equation}
Taking fourier's transform with respect to $x$ and $y$, we get
\begin{equation}
-D_pK^2\bar{G}+D_p\frac{\partial^2 \bar{G}}{\partial z^2}-2hnc\sigma_p\delta(z)\bar{G}=-e^{ik_xx^\prime+ik_yy^\prime}\delta(z)
\end{equation}
where $K^2=k_x^2+k_y^2$ and 
\begin{equation}
\bar{G}(k_x,x^\prime,k_y,y^\prime,z,z^\prime)=\int_{-\infty}^\infty dx\int_{-\infty}^\infty dy G(x,x^\prime,y,y^\prime,z,z^\prime)e^{ik_xx+ik_yy}
\end{equation}
Also note that we have assigned $z^\prime=0$ in Eq. (B2). Solving Eq. (B2) for the regions $z>0$ and $z<0$ by using the boundary conditions at $z=\pm H$, we get
\begin{equation}
\bar{G}(k_x,x^\prime,k_y,y^\prime,z)=\bar{G}_0(k_x,x^\prime,k_y,y^\prime,0)\frac{\mathrm{sinh}\left[K(H-|z|)\right]}{\mathrm{sinh}(KH)}
\end{equation}  
The continuity equation at $z=0$ is obtained by integrating Eq. (B2) over $z$ around $0$ as
\begin{equation}
D_p\left[\frac{\partial \bar{G}}{\partial z}\right]^{+0}_{-0}-2hnc\sigma_p\bar{G}_0+e^{ik_xx^\prime+ik_yy^\prime}=0
\end{equation}
Solving for $\bar{G}_0$ from Eqs. (B4 $\&$ B5), we get
\begin{equation}
\bar{G}_0(k_x,x^\prime,k_y,y^\prime,0)=\frac{e^{ik_xx^\prime+ik_yy^\prime}}{\left[2D_pK\mathrm{coth}(KH)+2hnc\sigma_p\right]}
\end{equation}
and substituting it back to Eq. (B4), we obtain  
\begin{equation}
\bar{G}(k_x,x^\prime,k_y,y^\prime,z)=\frac{e^{ik_xx^\prime+ik_yy^\prime}}{\left[2D_pK\mathrm{coth}(KH)+2hnc\sigma_p\right]}\times\frac{\mathrm{sinh}\left[K(H-|z|)\right]}{\mathrm{sinh}(KH)}
\end{equation}
Taking an inverse fourier's transform of Eq. (B7) gives
\begin{equation}
G(x,x^\prime,y,y^\prime,z)=\frac{1}{4\pi^2}\int_{-\infty}^\infty dk_x\int_{-\infty}^\infty dk_y \frac{e^{-ik_x(x-x^\prime)-ik_y(y-y^\prime)}}{\left[2D_pK\mathrm{coth}(KH)+2hnc\sigma_p\right]}\times \frac{\mathrm{sinh}\left[K(H-|z|)\right]}{\mathrm{sinh}(KH)}
\end{equation}
Eq. (B8) is a two dimensional fourier transform which can be easily simplified in the form of Hankel transform by changing the variables as
\begin{equation}
k_x=K\mathrm{cos}\phi\quad;\quad	k_y=K\mathrm{sin}\phi \quad;\quad x-x^\prime=(r-r^\prime)\mathrm{cos}\theta\quad;\quad	y-y^\prime=(r-r^\prime)\mathrm{sin}\theta
\end{equation}  
Then,
\begin{equation}
G(r,r^\prime,z)=\frac{1}{4\pi D_p}\int^\infty_0\frac{\mathrm{sinh}[K(H-|z|)]}{\mathrm{sinh}(KH)\left[K\mathrm{coth}(KH)+\frac{2hnc\sigma_p}{2D_p}\right]}\times \mathrm{J_0}\left[K(r-r^\prime)\right]KdK
\end{equation}
where $\mathrm{J_0}$ is the Bessel function of order $0$. Having calculated the Green's function, the CR density at $r=0$ due to a uniform source distribution $S_p(\textbf{r}^\prime,E)=\Re Q_p(E)\delta(z^\prime)$ in the Galactic plane with radial distances between  $r_1$ and $r_2$ from the Galactic center is given by
\begin{equation}
N_p(z,E)=2\pi\Re\int_{r_1}^{r_2} r^\prime dr^\prime G(r=0,r^\prime,z)Q_p(E)
\end{equation} 
This gives,
\begin{equation}
N_p(z,E)=\frac{\Re Q_p(E)}{2D_p}\int_0^\infty\frac{\mathrm{sinh}\left[K(H-|z|)\right]}{\mathrm{sinh}(KH)\left[K\mathrm{coth}(KH)+\frac{2hnc\sigma_p}{2D_p}\right]}\times \left[r_2\mathrm{J_1}(Kr_2)-r_1\mathrm{J_1}(Kr_1)\right]dK
\end{equation} 
where $\mathrm{J_1}$ represents the Bessel function of order $1$ and we have used the standard relation
\begin{equation}
\int_{r_1}^{r_2}r^\prime \mathrm{J_0}(Kr^\prime)dr^\prime=\frac{1}{K}\left[r_2\mathrm{J_1}(Kr_2)-r_1\mathrm{J_1}(Kr_1)\right]
\end{equation}
By setting $r_1=0$ and $r_2=R$, Eq. (B12) can be used to find the CR primary density due to all the sources extended upto a radial distance $R$ from the Earth.  

\end{document}